\documentclass[twocolumn]{aastex63}
\usepackage{amsmath}	
\usepackage{amssymb}	
\usepackage{graphicx}

\newcommand{\kms}{km~s$^{-1}$}

\newcommand{\unitlum}{erg~s$^{-1}~$}

\newcommand{\ha}{H$\alpha$}
\newcommand{\hb}{H$\beta$}
\newcommand{\hg}{H$\gamma$}
\newcommand{\hd}{H$\delta$}

\newcommand{\sii}{[\ion{S}{2}]}
\newcommand{\niii}{\ion{N}{3}}
\newcommand{\oiii}{[\ion{O}{3}]}
\newcommand{\heii}{\ion{He}{2}}
\newcommand{\hei}{\ion{He}{1}}

\newcommand{\tad}{F01004}

\accepted{January 22, 2021}
\submitjournal{ApJ}

\shorttitle{Evolution of the He lines in F01004}
\shortauthors{Cannizzaro et al.}

\graphicspath{{./}{figures/}}

\begin{document}

\title{Spectroscopic monitoring of the candidate tidal disruption event in F01004$-$2237}

\correspondingauthor{Giacomo Cannizzaro}
\email{g.cannizzaro@sron.nl}

\author[0000-0003-3623-4987]{Giacomo Cannizzaro}
\affiliation{SRON, Netherlands Institute for Space Research\\
Sorbonnelaan, 2, 3584CA Utrecht, the Netherlands}
\affiliation{Department of Astrophysics/IMAPP, Radboud University\\
P.O. Box 9010, 6500 GL Nijmegen, the Netherlands}

\author[0000-0001-5679-0695]{Peter G. Jonker}
\affiliation{Department of Astrophysics/IMAPP, Radboud University\\
P.O. Box 9010, 6500 GL Nijmegen, the Netherlands}
\affiliation{SRON, Netherlands Institute for Space Research\\
Sorbonnelaan, 2, 3584CA Utrecht, the Netherlands}

\author[0000-0003-0245-9424]{Daniel Mata-S\'anchez}
\affiliation{Jodrell Bank Centre for Astrophysics, School of Physics and Astronomy,\\ The University of Manchester, Manchester, M13 9PL, UK}

\begin{abstract}

We present results of spectroscopic monitoring observations of the Ultra-Luminous Infra Red Galaxy F01004$-$2237. This galaxy was observed to undergo changes in its optical spectrum, detected by comparing a spectrum from 2015 with one from 2000. These changes were coincident with photometric brightening. The main changes detected in the optical spectrum are enhanced \heii\ $\lambda$4686 emission and the appearance of \hei\ $\lambda$3898,$\lambda$5876 emission lines. The favoured interpretation of these changes was that of a tidal disruption event (TDE) happening in 2010. However, subsequent work suggested that these changes are caused by another hitherto unknown reason related to variations in the accretion rate in the active galactic nucleus (AGN). Our optical spectroscopic monitoring observations show that the evolution of the He lines is in line with the evolution seen in TDEs and opposite of what observed from reverberation mapping studies of AGNs, renewing the discussion on the interpretation of the flare as a TDE.

\end{abstract}


\section{Introduction} \label{sec:intro}
A star traveling through the nuclear region of a galaxy can find itself so close to the central super-massive black hole (SMBH) that it will be ripped apart by the tidal forces of the BH \citep{hills75,rees88, evans89}. During this tidal disruption event (TDE), part of the stellar material will be bound to the SMBH, ultimately accreting onto it and giving rise to a luminous flare.
Optical spectroscopy of TDEs shows a large degree of heterogeneity in terms of presence/absence of emission lines and their observed properties. In general, they are characterised by a blue continuum, broad H and He emission lines, luminosities of order $10^{44}$ \unitlum and typical evolution timescales of months up to a year \citep[see][for a review]{vanvelzen20b}, but there are examples of more long-lived TDEs \citep{lin17,mattila18} in other wavebands.

TDEs are an important tool to detect dormant SMBHs and the majority of these events are found in otherwise inactive galaxies. Study of TDEs in galaxies that host an Active Galactic Nucleus (AGN) is hindered by the intrinsic difficulty in distinguishing a TDE from emission from the AGN. Nonetheless, TDEs have been found in low luminosity AGNs \citep[e.g.][]{prieto16,onori19,nicholl20} and they have been invoked to explain extreme variability in AGNs \citep{merloni15,graham16,cannizzaro20}.

\citet{tadhunter17} (from here on, T17) presents the serendipitous discovery of spectral changes in the Ultra-Luminous Infra Red Galaxy (ULIRG, characterised by strong star formation and accretion onto the central SMBH, due to recent mergers) F01004$-$2237 (from here on, \tad) at z=0.118 that hosts an AGN and an SMBH with a mass $M_{bh} \approx 2.5\times10^7$ M$_\odot$ \citep{dasyra06}. Comparing an optical spectrum from September 2015 with one from February 2000 one finds that prominent spectral changes are; enhanced \heii\ $\lambda$4868 line emission; and the appearance of \hei\ lines at $\lambda$3898,$\lambda$5876. The historical lightcurve from the Catalina Sky Survey (CSS, \citealt{Drake09}) shows a clear brightening starting around 2010. This together with the spectroscopic changes led T17 to propose a TDE, triggered in 2010, as the explanation for the observed changes.

Here, we report results of optical spectroscopic monitoring observations of \tad\, obtained over the period August 2017 -- September 2020. We discuss the evolution of the broad He emission lines, which become narrower and fainter over time. This is typical behaviour among TDEs, and contrary to that observed in reverberation mapping studies of AGNs \citep{peterson04}.

\section{Observations and data reduction}
\subsection{Spectroscopic data}
Optical spectra of \tad\ were acquired with the Intermediate dispersion Spectrograph and Imaging System (ISIS) and the Auxiliary-port CAMera (ACAM) spectrographs, mounted at the Cassegrain focus of the William Herschel Telescope (WHT), and the Device Optimized for the LOw RESolution (DOLORES), installed at the Nasmyth B focus of the Telescopio Nazionale Galileo (TNG). Both telescopes are part of the Roque de los Muchachos observatory (La Palma, Spain). We also re-analysed the ISIS spectrum originally reported in T17 and we add two epochs of observations retrieved from the WHT archive\footnote{\url{http://casu.ast.cam.ac.uk/casuadc/ingarch/query}}, to the sample of spectra we report on here. Overall, these spectra were taken over a period spanning September 2015 - 2020. In the case of ISIS, different grisms with different resolutions (R300B and R600B for the blue arm and R158R, R316R and R600R for the red arm) were used, while ACAM was always used in combination with the V400 grism and the GG395 order-sorting filter and DOLORES was used with the LR-B grism. 

\begin{figure*}
        \includegraphics[width=2\columnwidth]{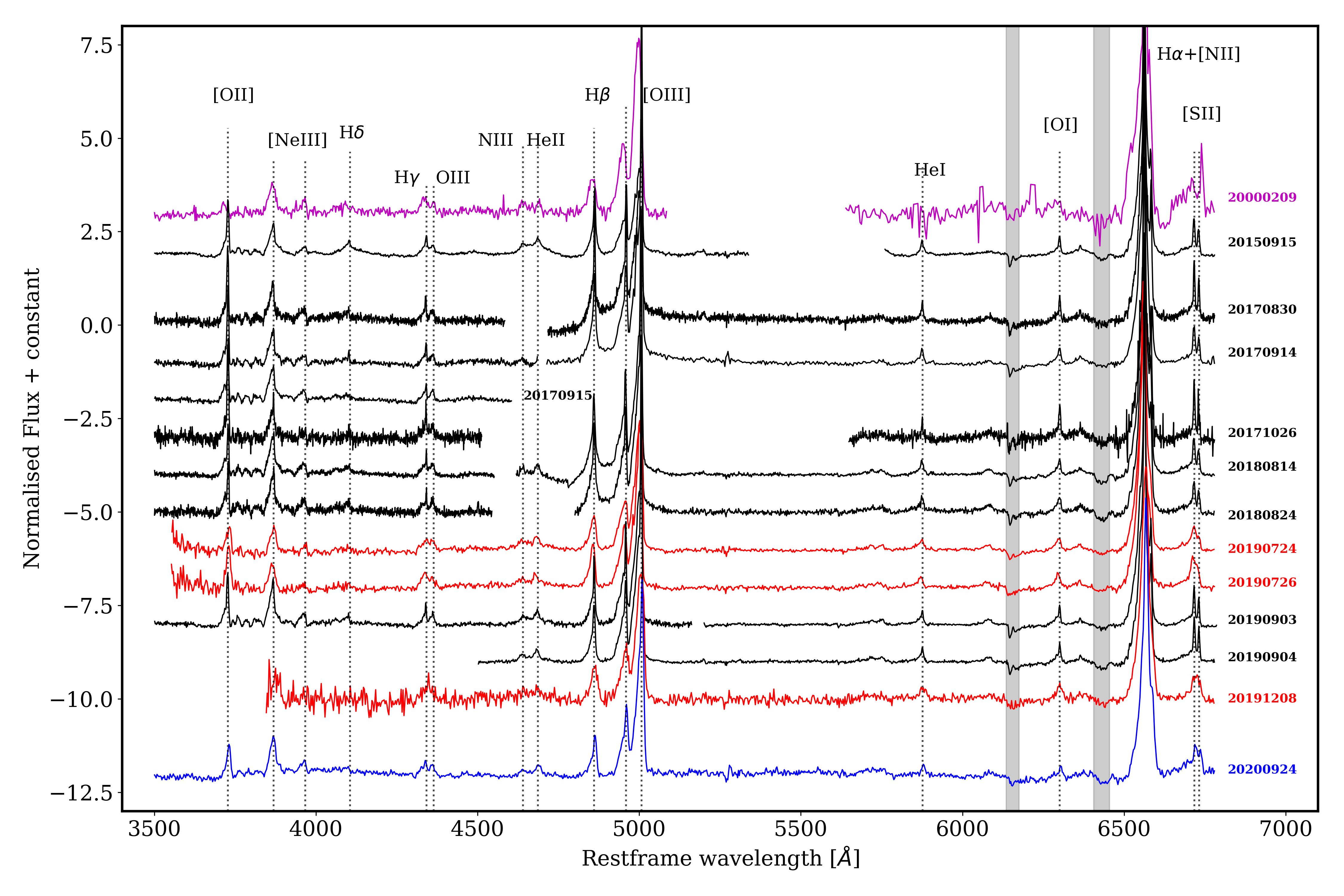} 
        \caption{The sequence of spectra taken with HST/STIS (magenta), WHT/ISIS (black), WHT/ACAM (red) and TNG/DOLORES (blue) under study in this paper. For each spectrum the date of observation is given on the right hand side. The dotted lines indicate the wavelength of the main emission lines. The grey bands indicate wavelength ranges affected by telluric absorption. The spectra are not flux calibrated and the continuum has been normalised (see text for details).}
        \label{fig:spectra}   
    \end{figure*}

Data were reduced using standard {\sc iraf} \citep{tody86} procedures for flat field and bias correction, and wavelength calibration with arc lamps. Cosmic-rays we removed using the {\sc lacosmic} procedure from \citet{lacosmic}. Standard star observations were not performed at all epochs and therefore the spectra are not flux calibrated. Instead, we normalise the spectra by dividing them by a polynomial fitted to the continuum (of order 3 to 5, depending on the spectrum), masking regions with prominent emission and absorption lines during the fit. We finally also include the spectrum taken on 2000 February 9 with the Space Telescope Imaging Spectrograph (STIS) on board of the Hubble Space Telescope (HST), retrieved from the online archive\footnote{\url{https://archive.stsci.edu/hst/}}. This spectrum and also the 2015 WHT/ISIS spectrum have also been presented by T17. All the observations we performed were carried out with the slit at parallactic angle, while observations retrieved from the online WHT archive were not. A journal of spectroscopic observations, with the resolution of the instrument/grism used is reported in Table~\ref{tab:specobs} and all the spectra are plotted in Fig.~\ref{fig:spectra}.

\begin{table*}
	\centering
	 \begin{center}
	\small
 	\caption{A log of the spectroscopic observations used in this paper.}

 	\label{tab:specobs}
 	\tabcolsep=0.11cm

	\begin{tabular}{lcccccccc}
		\hline
		MJD$^{(1)}$  & UTC Date  & {Telescope-instrument} & Gratings$^{(2)}$ & exposure time$^{(2)}$   & slit width  & seeing$^{(3)}$ & $\rm \Delta\lambda_{arc}$ $^{(4)}$ & $\rm  \Delta\lambda^{(5)}_{corr}$ 	 \\
        	{[days]} & 	 & &  blue, red   &    [s]             &[$''$] &[$''$]  &[\AA] &[\AA]   \\
        \hline
        51583.73$*$ & 2000/02/09 & HST-STIS & G430L, G750L  &	2904, 1754 $^{(6)}$ & 0.2 & -   & 4.1, 7.4 & 4.1, 7.4   \\
        57280.08$*$ & 2015/09/15 & WHT-ISIS & R300B, R316R  &	6x1000, 6x1000	& 1.5     & 0.60 & 4.1, 5.1 & 1.6, 2.0  \\ 
        57996.19    & 2017/08/30 & WHT-ISIS & R300B, R316R  &   2x1800, 2x1800	& 1.0     & 0.80 & 3.7, 3.8 & 3.0, 3.0  \\
        58011.00    & 2017/09/14 & WHT-ISIS & R300B, R158R  &	1800, 1800	    & 1.0     & 0.90 & 3.9, 3.4 & 3.5, 3.1  \\
        58012.12    & 2017/09/15 & WHT-ISIS & R300B		    &   2x1800 		    & 1.0     & 0.60 & 3.9      & 2.3       \\
        58052.99    & 2017/10/26 & WHT-ISIS & R600B, R600R  &	2x1800, 2x1800	& 1.5     & 2.50 & 3.0, 2.6 & 3.0, 2.6  \\
        58345.15    & 2018/08/14 & WHT-ISIS & R300B, R158R  &	2x1800, 2x1800	& 1.0     & 0.35 & 3.7, 7.0 & 1.3, 2.4  \\
        58355.22    & 2018/08/24 & WHT-ISIS & R300B, R158R  &	1800, 1800	    & 1.0     & 0.40 & 3.7, 7.0 & 1.5, 2.8  \\
        58689.20    & 2019/07/24 & WHT-ACAM & V400	        &   1800		    & 1.0     & 0.65 & 13.6 & 8.9       \\
        58691.20    & 2019/07/26 & WHT-ACAM & V400	        &   1800		    & 1.0     & 0.75 & 13.6 & 10.3      \\
        58730.05$*$ & 2019/09/03 & WHT-ISIS & R300B, R316R  &	3x900, 6x1200	& 1.3     & 0.60 & 4.9, 4.4 & 2.3, 2.0  \\
        58731.13$*$ & 2019/09/04 & WHT-ISIS & R316R		    &   7x1200	        & 1.3     & 0.60 & 4.4 & 2.0       \\
        58825.82    & 2019/12/08 & WHT-ACAM & V400	        &  1800		        & 1.5     & 1.10 & 19 & 14.1      \\
        59117.10    & 2020/09/24 & TNG-DOLORES & LR-B		    &  2x1800		    & 1.0     & 0.75 & 7.3 & 5.5       \\

		\hline
\end{tabular}

\textit{Note.}(1) Modified Julian Day of observations. (2) In the case of ISIS and STIS, the blue and red arms have different gratings, with different resolutions. The ISIS grating names ending with B denote those used in the blue arm, the ones ending in R those of the red arm. The exposure time is given for each grating separately. Multiple exposures obtained on the same day have been averaged after extraction. (3) The reported atmospheric seeing is the average value over the total exposure time. (4,5) Resolution element, as measured from arc lines and corrected for the seeing when this is lower than the slit width, by multiplying $\Delta\lambda_{arc}$ by the ratio between seeing and slit width. For ISIS and STIS, this is given for each grating. (6) The total exposure time of the STIS spectra is given (720+720+780+754s for G430L and 624+624+506 for G750L). Epochs marked with $*$ were retrieved from the respective observatory data archives (see text).
\end{center}
\end{table*}

\section{Data analysis and results}
\label{sec:data}
The spectra (see Fig.~\ref{fig:spectra}) reveal a number of emission lines from different atomic species: [\ion{O}{2}] $\lambda$3727, [\ion{Ne}{3}] $\lambda$3869, \hd, \hg, \oiii\ $\lambda$4363,$\lambda$4959,$\lambda$5007, \niii\ $\lambda$4640, \heii\ $\lambda$4686, \hb, \hei\ $\lambda$3889,$\lambda$5876, [\ion{O}{1}] $\lambda$6300, \ha, [\ion{N}{2}] $\lambda$6548,$\lambda$6584 and \sii\ $\lambda$6717,6731. Comparing the STIS spectrum (2000) with the ISIS spectrum obtained in 2015 shows that He lines were either not detected (\hei) or much fainter (\heii) in the former. 

\begin{figure}
        \includegraphics[width=\columnwidth]{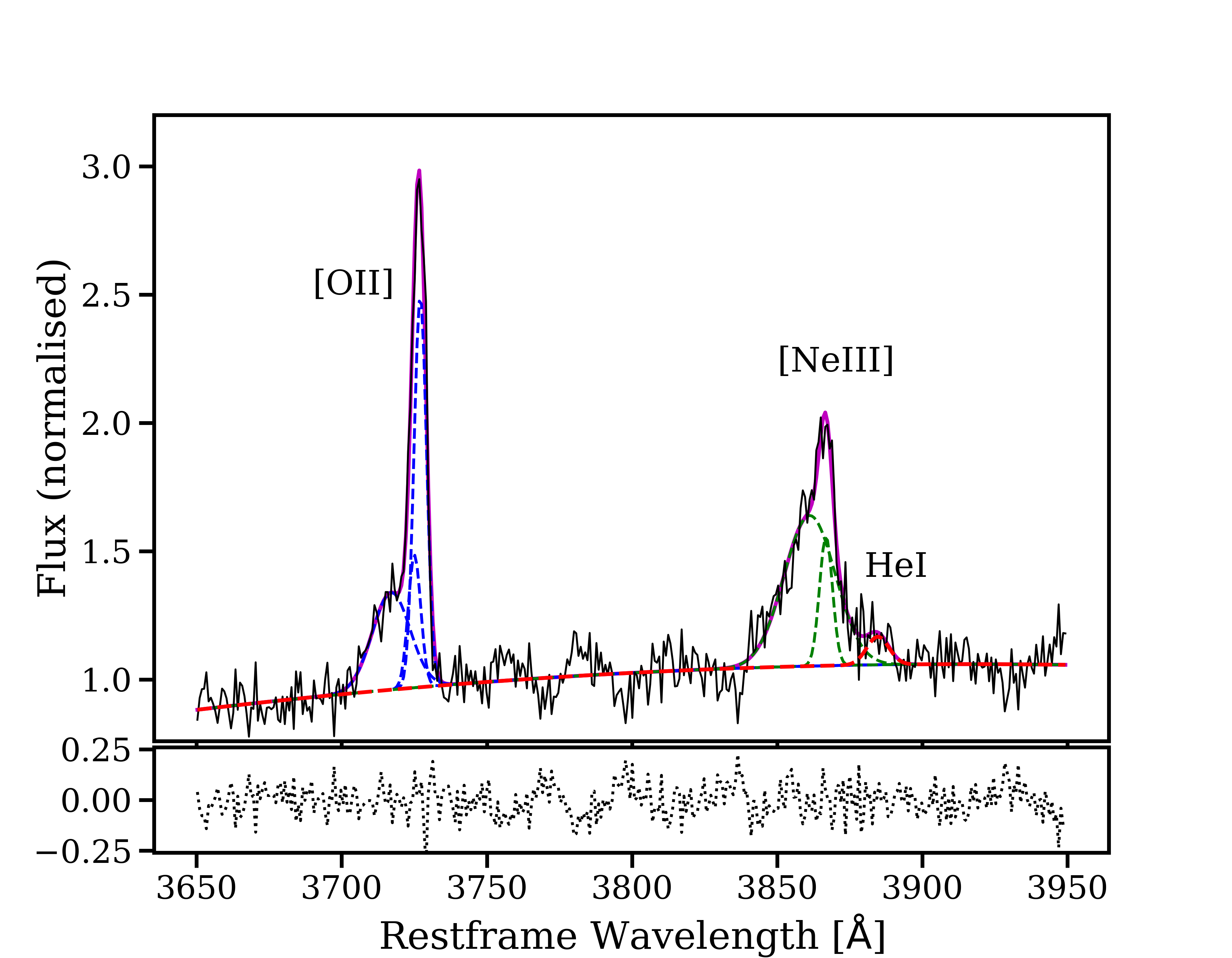} 
        \caption{An example fit to the [\ion{O}{2}], the [\ion{Ne}{3}] and the \hei$\lambda$3889 emission lines, in the ISIS spectrum of 2017 Aug 30. The dashed lines indicate the different Gaussians (blue for [\ion{O}{2}], green for [\ion{Ne}{3}] and red for \hei$\lambda$3889) used to describe the various emission lines and their components. The solid magenta line is the best-fit fit function. The residuals of the fit are shown in the bottom panel.}
        \label{fig:hbfit}   
    \end{figure}

The \niii\ $\lambda$4640 emission line is consistent with being caused by Wolf-Rayet stars (see T17 and references therein), as is the \heii\ $\lambda$4686 in the pre-flare HST spectrum. The rest of the emission lines are typically observed in AGNs. We fit the emission lines with a combination of Gaussian functions and a polynomial (to fit the local continuum), using the \textsc{python} package \textsc{lmfit} \citep{lmfit}. We plot an example of the fit to the \hb\ and \oiii\ $\lambda$4959,$\lambda$5007 emission lines in Fig. \ref{fig:hbfit}. All emission lines caused by activity in the host galaxy show the same structure, with a narrow peak and a broader, blue-shifted, base. There is evidence for an additional, third, component, with Full Width Half Maximum (FWHM) in between that of the narrow peak and that of the broader base. The central wavelength also between that of the broad and that of the narrow component. We were able to constrain this third component only for the strongest emission lines when detected with the highest signal--to--noise ratios (SNRs). 

In the lower resolution spectra as well as in lower SNR spectra, the \hei\ $\lambda$5876 emission is often well-fit by a single Gaussian function, whereas in the higher resolution and the higher SNR spectra, we find that two Gaussians (a broad base and a narrower peak) are required to describe the emission line. The \heii\ $\lambda$4868 and \hei\ $\lambda$3889 emission lines are well fit by a single Gaussian.

During the fitting procedure, we forced the FWHM of the narrow lines close in wavelength to be the same. Furthermore, the wavelength separation of known line doublets has been fixed to their laboratory values. Both these actions served to reduce the number of degrees of freedom in the fit. The resulting values for the emission line parameters have been corrected for the instrumental broadening, using the $\rm\Delta\lambda_{corr}$ values reported in Table~\ref{tab:specobs}.
When the atmospherical seeing was smaller than the width of the slit, we calculate the instrumental resolution by multiplying the resolution element (measured from the arc lines) by the ratio between seeing and slit width. Due to the extended nature of the source, this may not be correct, as the resolution for lines emitted in the extended regions of the galaxy will depend on the geometry of those regions. As explained in Sec.~\ref{sec:disc}, we propose that the He emission lines (i.e., the emission lines we are interested in) are mostly emitted in the nuclear region. We can therefore assume that the emitting region of the He lines is point-like and our calculation of the spectral resolution a valid approximation.

The position of the galaxy on a Baldwin, Phillips \& Terlevich (BPT) diagram \citep{baldwin81} is in the ``composite'' region, between \ion{H}{2} regions and AGNs, suggesting that the galaxy is showing emission from both an AGN and star forming activity. This dual ionising source was already inferred by T17, where they also reported that the emission lines can have three components. We find that the FWHM of the narrow peaks varies somewhat with time. This can be explained if (part of) the narrow emission lines is caused by star formation from an extended, spatially resolved region of the galaxy. Then, the different slit widths, seeing conditions, and position angles of the slit contribute to us capturing a variable amount of light from the extended star-forming regions of the galaxy, causing the fraction of the emission line caused by star formation to vary. Furthermore, it is possible that the projected velocity of that component in the slit varies with our different spectra as well. 

\begin{figure*}
        \includegraphics[width=18cm]{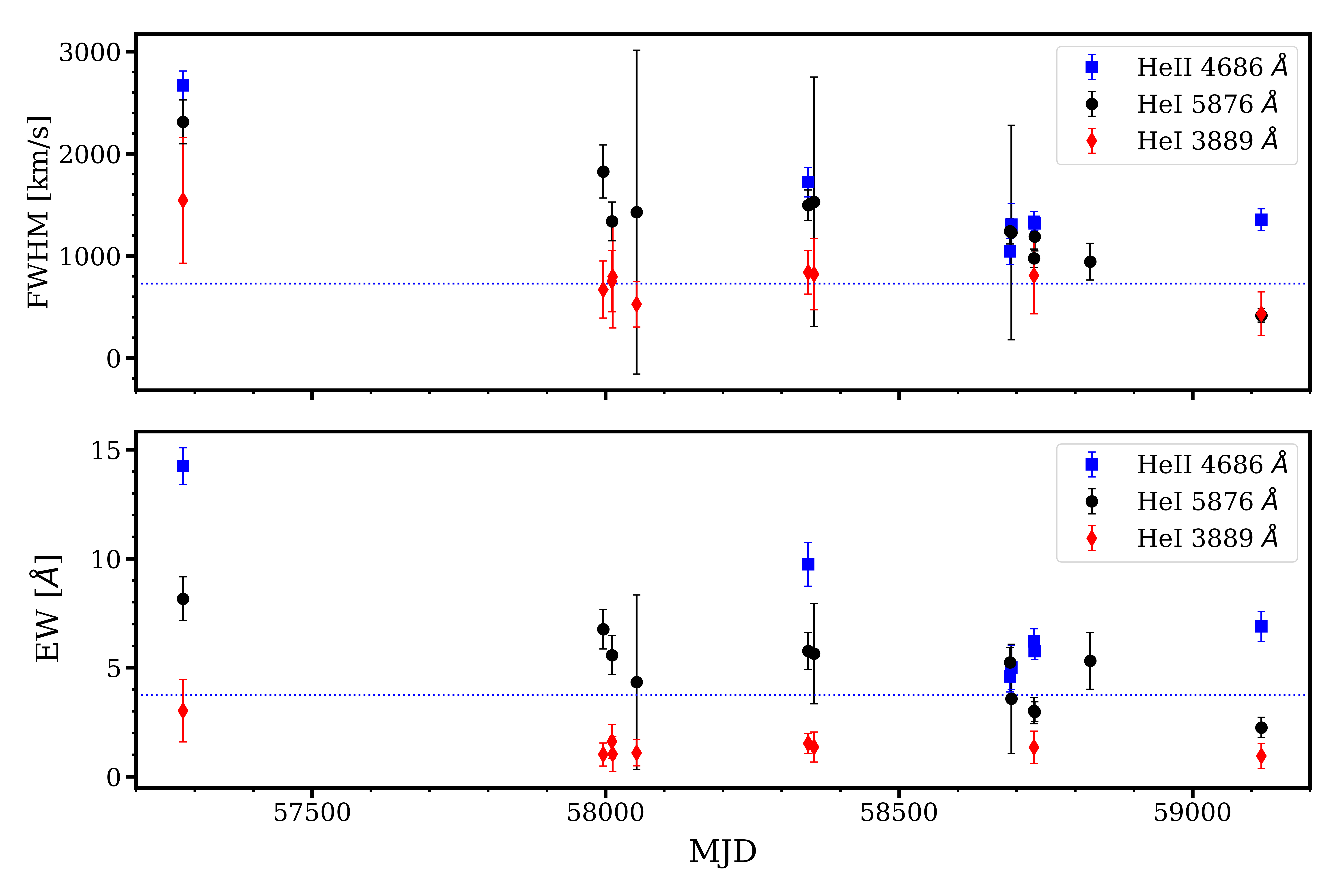} 
        \caption{Results of the fits to the emission lines of He in our WHT and TGN spectra. Top panel: FWHM of \heii\ $\lambda$4868 (blue squares), \hei\ $\lambda$5876 (black circles) and \hei\ $\lambda$3889 (red diamonds). Bottom panel: EW of the same emission lines (with the same color/marker combination as the top panel). The dotted blue lines indicate the value of the \heii\ line measured from the STIS spectra of 09 February 2000 (MJD 51584). The \hei\ lines were not detected in that spectrum. On the X-axis, the Modified Julian Date of the observations.}
        \label{fig:he_results}   
    \end{figure*}

We focus our analysis on the He lines, as T17 reported the largest variations in the properties of these lines. In Fig.~\ref{fig:he_results}, we plot the evolution of the equivalent width (EW) and FWHM of the He emission lines. In order to trace the evolution of the \hei\ $\lambda$5876 emission line, we combine the results of the two components, when resolved, by adding the EW and adding in quadrature the FWHM. The uncertainties were also added in quadrature for both the EW and the FWHM. We check the effect of this combination by comparing the values of the FWHM and EW for both components, when resolved, before and after their combination. The value of the FWHM is dominated by the broad component. The value of the EW of the narrow component does not show variations above the measurement error and therefore the effect of the combination is a shift in the vertical direction. The average value of the EW of the narrow component is $\approx$2.3 \AA.
The lack of evolution of the narrow \hei\ $\lambda$5876 suggests that this component is due to star formation and is not related to the transient event. We tried to check this by investigating if a \hei\ $\lambda$5876 emission line would be detected in the pre-flare HST-STIS spectrum. To do this, we added a Gaussian curve at the resolution of the instrument (see Tab.~\ref{tab:specobs}) and with EW$\approx$2.3 \AA, to the spectrum. Unfortunately, the area around 5876 \AA\ has a low SNR (see Fig.~\ref{fig:spectra}) and such an emission line could not be detected above the continuum noise in the HST spectrum.

The FWHM and EW of the \heii\ line show a gradual decrease over the duration of our follow-up campaign: the FWHM decreases from $\approx$2700 \kms to $\approx$1350 \kms and the EW from $\approx$14 \AA\ to $\approx$7 \AA. Both the FWHM and the EW have not yet returned to the values measured during the STIS spectrum. The evolution in the EW and FWHM of the \hei\ lines is less clear, but especially that of the 5876 line seems to follow a similar trend as that of \heii. In contrast, the other lines of similar width i.e.~the broad bases of the other emission lines, do not show significant evolution with time in their FWHM, while the EW of some of them (mainly \hb\ and \oiii$\lambda$4959,$\lambda$5007) show an increase between MJD 57280 and 58345, to subsequently decay back to the initial value (see Table \ref{tab:linefit}).

We measure the offset of the lines with respect to their rest-frame wavelength. It is important to note that in good observing conditions (i.e.,~when the value of the seeing in arcseconds is smaller than the slit width, which is the case for many of our observations, see Table \ref{tab:specobs}) the wavelength calibration of the source spectrum could be slightly shifted with respect to that derived by the arc lines (as the latter do fill the whole slit), due to possible imperfections in centering of the source in the slit. This can in turn affect the velocity offset measurements by a few times 100 \kms. However, 
the narrow lines do not show a significant offset with respect to their restframe wavelength (its absolute value is always below 200 \kms), implying that the shifts in the wavelength scale caused by the potential imperfect centring of the source in the slit are smaller than a few hundred \kms. This is in line with the fact that the extended nature of the source acts as a mitigating factor for any small imperfect centring of the source in the slit. 

The broad bases of the emission lines are all blue-shifted. We measure the shift of the broad bases with respect to the central wavelength of the narrow lines, assuming that those are at their restframe wavelength. Each broad line has a different blueshift, but none of them show a clear evolution with time. The velocity offset values are all $\lesssim10^3$ \kms.

For the \hei\ $\lambda$5876 lines, we only separate the narrow and broader component of the emission line when the higher resolution gratings were used. Only in those cases can we reliably check for a potential shift between the two components. Therefore, for the other He lines as well as for the \hei\ $\lambda$5876 lines, we measure both the shift with respect to the restframe wavelength and with respect to the central wavelength of the closest narrow line (\ion{Ne}{3} for \hei\ $\lambda$3889, \oiii\ $\lambda$5007 for \heii\ and \sii\ $\lambda$6731 for \hei\ $\lambda5876$). We find that the blueshift is a few 100 \kms and the two measurements are consistent with being the same, within the 1$\sigma$ uncertainties. Again, no clear evolution in time was present and at some epochs, the shift is consistent with zero.

\section{Discussion}
\label{sec:disc}
Our optical spectroscopic observations of F01004 reveal a gradual decrease in the FWHM and EW of the \heii\ and the \hei\ line at $\lambda5876$. The error bars on the individual measurements for the \hei\ $\lambda$3889 line are larger, especially those on the FWHM, making it more difficult to detect any trend in the FWHM for this line. We find no strong evolution in the properties of the other detected emission lines.

In T17, the authors proposed that the flare and the spectral changes can be explained by a TDE happening $\sim$5 years prior the 2015 ISIS spectrum. Our spectroscopic monitoring provides evidence in support of this interpretation. We first discuss this supportive evidence, before comparing our results with those from \citet{trakhtenbrot19} who come to the conclusion that F01004 and a group of sources showing similar spectroscopic and photometric characteristics (i.e.~a moderately broad \heii\ and a flare in the optical light curve) is caused by new, so far unexplained AGN variability.

The presence of broad He emission lines is common in TDEs and especially a broad \heii\ $\lambda$4686 emission is considered a strong indicator of such phenomena (see \citealt[e.g.~][]{vanvelzen20b}). However, while the FWHM of the \heii\ $\lambda$4686 emission line in \tad\, can be called ``broad'', in most TDEs the value of the FWHM is significantly larger than that observed in F01004 (cf.~\citealt{trakhtenbrot19}).
Nevertheless, as we showed in this paper, the evolution of its FWHM over time in F01004 is in line with what seen in other TDE candidates: the lines become narrower with time \citep{Holoien2014,brown17,onori19} as the flare decays and the line EW becomes lower. This behaviour is opposite of that seen in AGNs, where reverberation mapping studies have shown that the lines become broader with decreasing source luminosity \citep[e.g.~][]{peterson04}. 
The evolution of the EW of the He lines follows a more shallow but similar decay.

As mentioned in Sec.~\ref{sec:data}, the different position angle and observing conditions of the slit at different epochs make us capture a varying amount of light from any spatially extended emission lines regions of \tad. In the pre-outburst spectrum, the \heii\ and \niii\ line complex is mostly caused by emission from Wolf-Rayet stars (T17). The variation observed in the \heii\ line could, perhaps, be (partially) caused by this effect and not by an intrinsic change in the line emission properties. Whereas it is probably somewhat contrived that the combination of slit width, seeing, and parallactic angle work together to cause a gradual change in the EW, it is even more difficult to envisage how this would lead to a gradual decay in FWHM together with the observed evolution of EW. Finally, the \niii\ line does not show the same evolution as the \heii, further disfavouring this scenario. Therefore, we conclude that the observed trends in the EW and FWHM of the \heii\ and the \hei\ line at $\lambda5876$ is caused by changes in the accretion flow around the central supermassive black hole.

If a TDE caused the flaring activity in \tad, it must be quite long-lived, as our last spectrum, $\sim$10 years after the inferred date of the TDE, still shows \heii\ emission above the pre-outburst level. Such a long-lived TDE is rare but not unprecedented. \citet{lin17} reported on a TDE with X-ray emission lasting for more than a decade at around the Eddington level, explaining it as a TDE where the circularisation of the debris is slow. Late time X-ray observations of optically selected TDEs have also shown that the TDE phenomenon can be long lasting \citep{Jonker2020}. However, these examples are for the X-ray emission. So far, there is no reported optical emission that lasts this long. Perhaps, the presence of an AGN in F01004 could explain the decade-long signatures of a TDE. TDEs in AGNs are not well explored, but the interaction between the debris stream and the pre-existing accretion disk can significantly modify the canonical picture of a $\sim$year long decay \citep{chan19}. 

The BH mass of $\rm \sim 2.5\times10^7M_\odot$, calculated in \citet{dasyra06}, is below the $\rm 10^8M_\odot$ BH mass limit, for a non-rotating BH, above which the tidal radius for a sun-like star is inside the event horizon (and its disruption therefore impossible). In \citet{dasyra06}, the authors also explore the limits of dynamical mass measurements in the case of ULIRGs, due to the typically recent merger activity of these galaxies, finding that for a post-merger object like \tad, the measurement is reliable. We stress that even taking the $\rm M_{bh}$ value as an order of magnitude estimate does not affect the interpretation of the transient event, since for a more massive star and/or a spinning BH, the BH mass limit for a TDE to be possible can increase up to $10^9M_\odot$ \citep{kesden12}.

\citet{trakhtenbrot19} argues that \tad\ and two other similar transients are due to unusual AGN activity rather than being triggered by the disruption of a star. Their main argument for this is the long duration of the flare and the FWHM of the \heii\ line, which is smaller than what commonly observed in TDEs \citep[e.g][]{Arcavi2014}. Furthermore, the He lines in \tad\ show a lower blue-shift than typically seen in TDEs, where the broad emission lines are often blue-shifted by several 1000's \kms \citep[e.g.][]{nicholl20}. It is important to note that optical TDEs are often selected on the width of their emission lines and, therefore, that a \heii\ line is detected narrower than what seen in TDEs may be a product of selection bias.

\citet{trakhtenbrot19} associates the \niii\ and \heii\ emission with the Bowen fluorescence mechanism \citep[BF,][]{bowen34,bowen35}, a cascade of transitions initially triggered by enhanced UV and \heii\ Ly$\alpha$ emission. In the case of \tad, the \niii\ and \heii\ emission pre-outburst are associated with WR stars and the \niii\ line emission is not enhanced during the outburst, unlike the \heii\ emission line, implying that the \heii\ line has a different origin in \tad\, than perhaps in several other sources in the paper of \citet{trakhtenbrot19}.

We also find a broad, blue-shifted component to the forbidden emission lines. In AGNs, broad emission lines typically come from the high-velocity broad line region (BLR; \citealt{peterson06}). The density in the BLR is so high that the forbidden lines are collisionally suppressed. In F01004, the forbidden and permitted (hydrogen) emission lines have a similar structure, with a narrow peak and a broad base (with tentative evidence for a third, intermediate component). Furthermore, there is no evidence for changes in the FWHM and EW of these lines with time. This suggests that all the broad components to the forbidden and permitted lines are due to outflows from large spatial scales induced by the circumnuclear starburst and/or by the AGN, as commonly observed in ULIRGs (which of the two is the main driver of the outflow is still under debate \citep{rupke05,rodriguez13}).

In \citet{trakhtenbrot19}, the authors associate the observed broad \heii\ emission line with the BLR, illuminated by the enhanced UV emission. Given that in the case of \tad, both the forbidden as well as permitted lines have a broad component likely caused by an outflow, we do not seem to detect lines coming from the BLR, which may indicate that the BLR is obscured from our line of sight.  This, together with the fact that the evolution of the FWHM of the \heii\ line is unlike that in reverberation mapping studies of AGN, challenges the AGN scenario for the observed photometric and spectroscopic changes. Perhaps, following \citet{roth18}, the TDE-induced lines originate in a more spherical photosphere, and the FWHM is determined in part by electron scattering. The observed decrease of FWHM may then trace a decrease in the ambient density. 

Overall, the evolution of the He lines and the difference between their evolution and the evolution of the other lines seems to be inconsistent with typical TDE behaviour but also with typical AGN behaviour. 
\section*{Data availability}
All data will be made available in a reproduction package uploaded to Zenodo.
\acknowledgements
DMS acknowledges support from the ERC under the European Union’s Horizon 2020 research and innovation programme (grant agreement no. 715051; Spiders). This paper makes use of data obtained from the Isaac Newton Group Archive which is maintained as part of the CASU Astronomical Data Centre at the Institute of Astronomy, Cambridge. We thank T. van Grunsven, D. Lena, F. Onori, K. Maguire, D. Eapacchen and S. Prentice for carrying out part of the observatons used in this paper. This paper includes data obtained with the William Herschel Telescope as well as observations made with the Italian Telescopio Nazionale Galileo (TNG) operated on the island of La Palma by the Isaac Newton Group of Telescopes and the Fundación Galileo Galilei of the INAF (Istituto Nazionale di Astrofisica) respectively, at the Spanish Observatorio del Roque de los Muchachos of the Instituto de Astrofisica de Canarias.

\appendix
\section{Line fitting results}

\begin{center}
\begin{table*}
\caption{Results of the line fitting of the most prominent broad lines.}

\label{tab:linefit}

\scalebox{0.8}{%
\hspace*{-3.cm}\begin{tabular}{l  ccc @{\hskip 5mm}  ccc @{\hskip 5mm} ccc }
\hline
& \multicolumn{3}{c}{[\ion{O}{2}]} &  \multicolumn{3}{c}{[\ion{Ne}{3}]} & \multicolumn{3}{c}{\ion{He}{1}}  \\
MJD & $\lambda$ [\AA] & FWHM [\kms] & EW [\AA] & $\lambda$ [\AA] & FWHM [\kms] & EW [\AA] & $\lambda$ [\AA] & FWHM [\kms] & EW [\AA]  \\
\hline

51583.73 &	3714.4 $\pm$ 1.0 &	1100 $\pm$ 200 & 4.3 $\pm$ 1.0 & 3864.2 $\pm$ 0.6 & 1900 $\pm$	110 & 18.3 $\pm$ 1.4 &      $\cdots$    &      $\cdots$     &      $\cdots$    \\                  
57280.08 &	3719.0 $\pm$ 1.8 &	1130 $\pm$ 250 & 4.9 $\pm$ 1.2 & 3862.2 $\pm$ 2.2 & 1590 $\pm$	220 & 11.0 $\pm$ 1.9 & 3885.1 $\pm$ 5.2	& 1540  $\pm$  610 &	3.0 $\pm$ 1.4  \\
57996.19 &	3717.2 $\pm$ 1.3 &	1160 $\pm$ 210 & 5.9 $\pm$ 1.2 & 3861.1 $\pm$ 0.7 & 1660 $\pm$	100 & 13.0 $\pm$ 1.2 & 3884.8 $\pm$ 1.5	& 670   $\pm$  280 &	1.0 $\pm$ 0.5  \\
58011.00 &	3718.1 $\pm$ 1.6 &	1160 $\pm$ 260 & 6.1 $\pm$ 1.5 & 3861.6 $\pm$ 1.1 & 1530 $\pm$	140 & 12.9 $\pm$ 1.5 & 3883.2 $\pm$ 1.8	& 750   $\pm$  300 &	1.6 $\pm$ 0.8  \\
58012.12 &	3717.2 $\pm$ 1.0 &	1220 $\pm$ 190 & 6.1 $\pm$ 1.1 & 3861.6 $\pm$ 1.3 & 1570 $\pm$	160 & 13.4 $\pm$ 1.6 & 3882.4 $\pm$ 3.3	& 790	$\pm$  500 &	1.0 $\pm$ 0.8  \\
58052.99 &	3718.5 $\pm$ 1.5 &	1080 $\pm$ 250 & 7.0 $\pm$ 1.7 & 3862.7 $\pm$ 0.6 & 1700 $\pm$	110 & 14.7 $\pm$ 1.3 & 3886.8 $\pm$ 1.2	& 520	$\pm$  220 &	1.1 $\pm$ 0.6  \\
58345.15 &	3718.5 $\pm$ 0.8 &	1200 $\pm$ 140 & 5.9 $\pm$ 0.8 & 3862.6 $\pm$ 0.5 & 1600 $\pm$	60  & 16.0 $\pm$ 0.8 & 3885.2 $\pm$ 1.2	& 840	$\pm$  210 &	1.5 $\pm$ 0.5  \\
58355.22 &	3718.9 $\pm$ 0.9 &	1090 $\pm$ 170 & 5.7 $\pm$ 1.0 & 3862.9 $\pm$ 0.8 & 1510 $\pm$	110 & 14.6 $\pm$ 1.2 & 3883.8 $\pm$ 2.1	& 820   $\pm$  350 &	1.3 $\pm$ 0.7  \\
58689.20 &	3724.0 $\pm$ 4.8 &	1100 $\pm$ 620 & 6.6 $\pm$ 4.1 & 3865.7 $\pm$ 1.4 & 1590 $\pm$	280 & 13.6 $\pm$ 4.0 &      $\cdots$    &      $\cdots$    &      $\cdots$    \\
58691.20 &      $\cdots$    &      $\cdots$    &      $\cdots$    &      $\cdots$    &      $\cdots$    &      $\cdots$    &      $\cdots$    &      $\cdots$    &                \\
58730.05 &	3717.4 $\pm$ 1.6 &	1100 $\pm$ 250 & 5.2 $\pm$ 1.3 & 3861.8 $\pm$ 0.9 & 1570 $\pm$	110 & 15.5 $\pm$ 1.5 & 3882.9 $\pm$ 2.3	& 810	 $\pm$  370 &	1.3 $\pm$ 0.7  \\
58731.13 &      $\cdots$    &      $\cdots$    &      $\cdots$    &      $\cdots$    &      $\cdots$    &      $\cdots$    &      $\cdots$    &      $\cdots$    &                \\
58825.82 &      $\cdots$    &      $\cdots$    &      $\cdots$    &      $\cdots$    &      $\cdots$    &      $\cdots$    &      $\cdots$    &      $\cdots$    &                \\
59117.10 &	3723.8 $\pm$ 9.7 &	1160 $\pm$ 900 & 4.7 $\pm$ 5.1 & 3865.3 $\pm$ 0.9 & 1560 $\pm$	90  & 16.6 $\pm$ 1.7 & 3887.6 $\pm$ 1.1	& 430	 $\pm$  210 &	0.9 $\pm$ 0.6  \\

\hline
& \multicolumn{3}{c}{\ion{N}{3}} & \multicolumn{3}{c}{\ion{He}{2}} & \multicolumn{3}{c}{\hb} \\
MJD & $\lambda$ [\AA] & FWHM [\kms] & EW [\AA] & $\lambda$ [\AA] & FWHM [\kms] & EW [\AA] & $\lambda$ [\AA] & FWHM [\kms] & EW [\AA]  \\
\hline
51583.73 & 4647.1 $\pm$ 3.3	& 2680 $\pm$ 640 & 8.8 $\pm$ 2.7   & 4689.8 $\pm$ 1.2 & 730  $\pm$ 180 & 3.7  $\pm$ 1.3  & 4848.1 $\pm$ 5.0	& 1690 $\pm$ 440 & 13.6 $\pm$ 5.2 \\  
57280.08 & 4641.9 $\pm$ 1.0	& 1990 $\pm$ 130 & 7.7 $\pm$ 0.6   & 4685.4 $\pm$ 0.8 & 2670 $\pm$ 140 & 14.2 $\pm$ 0.8  & 4858.5 $\pm$ 0.3	& 1490 $\pm$ 50	 & 18.5 $\pm$ 0.9 \\
57996.19 &      $\cdots$    &      $\cdots$    &      $\cdots$    &      $\cdots$    &      $\cdots$    &      $\cdots$    & 4857.6 $\pm$ 0.5	& 2030 $\pm$ 80	 & 30.2 $\pm$ 1.7 \\
58011.00 &      $\cdots$    &      $\cdots$    &      $\cdots$    &      $\cdots$    &      $\cdots$    &      $\cdots$    & 4857.0 $\pm$ 1.0	& 1840 $\pm$ 160 & 20.8 $\pm$ 2.6 \\
58012.12 &      $\cdots$    &      $\cdots$    &      $\cdots$    &      $\cdots$    &      $\cdots$    &      $\cdots$    &      $\cdots$    &      $\cdots$    &      $\cdots$    \\
58052.99 &      $\cdots$    &      $\cdots$    &      $\cdots$    &      $\cdots$    &      $\cdots$    &      $\cdots$    &      $\cdots$    &      $\cdots$    &      $\cdots$    \\
58345.15 & 4641.9 $\pm$ 1.8	& 2000 $\pm$ 340 & 7.6 $\pm$ 1.6   & 4685.2 $\pm$ 0.9 & 1720 $\pm$ 140 & 9.7 $\pm$ 1.0   & 4857.1 $\pm$ 0.6	& 2100 $\pm$ 110 & 28.0 $\pm$ 2.0 \\
58355.22 &      $\cdots$    &      $\cdots$    &      $\cdots$    &      $\cdots$    &      $\cdots$    &      $\cdots$    & 4856.6 $\pm$ 0.7	& 1940 $\pm$ 120 & 25.3 $\pm$ 2.2 \\
58689.20 & 4644.5 $\pm$ 1.9	& 2430 $\pm$ 370 & 6.4 $\pm$ 1.2   & 4685.6 $\pm$ 0.8 & 1040 $\pm$ 120 & 4.6 $\pm$ 0.7   & 4854.3 $\pm$ 0.8	& 1410 $\pm$ 90	 & 15.1 $\pm$ 1.5 \\
58691.20 & 4640.1 $\pm$ 2.2	& 1910 $\pm$ 410 & 4.7 $\pm$ 1.3   & 4682.8 $\pm$ 1.3 & 1300 $\pm$ 210 & 5.0 $\pm$ 1.0   & 4849.6 $\pm$ 2.8	& 1330 $\pm$ 270 & 9.1  $\pm$ 2.4 \\
58730.05 & 4645.1 $\pm$ 1.4	& 2530 $\pm$ 260 & 7.5 $\pm$ 0.9   & 4684.5 $\pm$ 0.7 & 1330 $\pm$ 100 & 6.2 $\pm$ 0.6   & 4854.4 $\pm$ 0.5	& 1320 $\pm$ 70	 & 13.8 $\pm$ 1.1 \\
58731.13 & 4641.2 $\pm$ 0.9	& 1950 $\pm$ 160 & 4.9 $\pm$ 0.5   & 4682.2 $\pm$ 0.4 & 1320 $\pm$ 70  & 5.7 $\pm$ 0.4   & 4853.0 $\pm$ 0.5	& 1460 $\pm$ 50	 & 14.2 $\pm$ 0.8 \\
58825.82 &      $\cdots$    &      $\cdots$    &      $\cdots$    &      $\cdots$    &      $\cdots$    &      $\cdots$    & 4854.5 $\pm$ 5.2	& 1650 $\pm$ 530 & 11.1 $\pm$ 6.2 \\
59117.10 & 4644.2 $\pm$ 1.5	& 2510 $\pm$ 292 & 7.5 $\pm$ 1.2   & 4688.0 $\pm$ 0.7 & 1353.1 $\pm$ 110 & 6.9 $\pm$ 0.7 & 4852.4 $\pm$ 1.3	& 1470 $\pm$ 140 & 10.2 $\pm$ 1.2 \\

\hline
& \multicolumn{3}{c}{[\ion{O}{3}]} & \multicolumn{3}{c}{\ion{He}{1}} & \multicolumn{3}{c}{\ha} \\
MJD & $\lambda$ [\AA] & FWHM [\kms] & EW [\AA] & $\lambda$ [\AA] & FWHM [\kms] & EW [\AA] & $\lambda$ [\AA] & FWHM [\kms] & EW [\AA]  \\
\hline
51583.73 & 4987.6 $\pm$ 2.7 & 1340 $\pm$ 230 & 69.7 $\pm$ 17    &      $\cdots$    &      $\cdots$     &      $\cdots$   & 6548.2 $\pm$ 2.6 & 2290 $\pm$ 180	& 180	 $\pm$ 18  \\
57280.08 & 4992.8 $\pm$ 0.5 & 1640 $\pm$ 40  & 43.9 $\pm$ 2.0   & 5880.2 $\pm$ 1.2 & * 2310 $\pm$ 210  & * 8.2 $\pm$ 1.0 & 6559.7 $\pm$ 0.2 & 1830 $\pm$ 20 	& 163.5	 $\pm$ 3.5   \\
57996.19 & 4993.3 $\pm$ 0.6 & 1600 $\pm$ 50  & 52.4 $\pm$ 2.9   & 5864.8 $\pm$ 2.3 & * 1820 $\pm$ 260  & * 6.8 $\pm$ 0.9 & 6559.1 $\pm$ 0.2 & 1770 $\pm$ 20 	& 174.1	 $\pm$ 2.8   \\
58011.00 & 4995.8 $\pm$ 0.4 & 1470 $\pm$ 40  & 63.9 $\pm$ 2.2   & 5867.4 $\pm$ 2.1 & * 1340 $\pm$ 190  & * 5.6 $\pm$ 0.9 & 6556.9 $\pm$ 0.4 & 1720 $\pm$ 40 	& 144.7	 $\pm$ 5.3   \\
58012.12 &      $\cdots$    &      $\cdots$    &      $\cdots$  &      $\cdots$    &      $\cdots$     &     $\cdots$    &      $\cdots$    &      $\cdots$    &      $\cdots$    \\
58052.99 &      $\cdots$    &      $\cdots$    &      $\cdots$  & 5872.7 $\pm$ 2.0 & * 1430 $\pm$ 1590 & * 4.3 $\pm$ 4.0 & 6559.8 $\pm$ 0.3 & 1740 $\pm$ 30 	& 174.7	 $\pm$ 4.5   \\
58345.15 & 4992.4 $\pm$ 0.5 & 1690 $\pm$ 40  & 64.2 $\pm$ 2.7   & 5867.8 $\pm$ 1.5 & * 1490 $\pm$ 150  & * 5.8 $\pm$ 0.8 & 6558.9 $\pm$ 0.3 & 1900 $\pm$ 30 	& 196.2	 $\pm$ 5.5   \\
58355.22 & 4992.0 $\pm$ 0.5 & 1660 $\pm$ 50  & 61.8 $\pm$ 3.1   & 5857.1 $\pm$ 5.1 & * 1530 $\pm$ 1220 & * 5.6 $\pm$ 2.3 & 6558.5 $\pm$ 0.3 & 1910 $\pm$ 30 	& 192.0	 $\pm$ 5.4   \\
58689.20 & 4991.0 $\pm$ 0.9 & 1600 $\pm$ 70  & 56.9 $\pm$ 4.4   & 5862.0 $\pm$ 5.4 &   1240 $\pm$ 120  &   5.2 $\pm$ 0.7 & 6556.3 $\pm$ 0.5 & 1870 $\pm$ 30 	& 166.1	 $\pm$ 5.9   \\
58691.20 & 4991.7 $\pm$ 1.2 & 1400 $\pm$ 120 & 52.1 $\pm$ 5.0   & 5856.3 $\pm$ 12.3& * 1230 $\pm$ 1050 & * 3.6 $\pm$ 2.5 & 6554.4 $\pm$ 0.5 & 1870 $\pm$ 40 	& 166.0	 $\pm$ 7.0   \\
58730.05 & 4991.7 $\pm$ 0.5 & 1630 $\pm$ 40  & 56.0 $\pm$ 2.8   & 5868.1 $\pm$ 0.8 & * 980  $\pm$ 90   & * 3.0 $\pm$ 0.6 & 6557.1 $\pm$ 0.2 & 1920 $\pm$ 20 	& 191.6	 $\pm$ 3.6   \\
58731.13 & 4990.9 $\pm$ 0.4 & 1670 $\pm$ 30  & 58.7 $\pm$ 2.0   & 5867.3 $\pm$ 1.4 & * 1190 $\pm$ 140  & * 3.0 $\pm$ 0.5 & 6558.5 $\pm$ 0.3 & 1850 $\pm$ 20 	& 179.2	 $\pm$ 4.0   \\
58825.82 & 4993.2 $\pm$ 3.7 & 1010 $\pm$ 470 & 36 $\pm$ 20      & 5877.7 $\pm$ 1.3 &   940  $\pm$ 180  &   5.3 $\pm$ 1.3 & 6555.0 $\pm$ 4.5 & 2040 $\pm$ 190	& 117	 $\pm$ 22   \\
59117.10 & 4995.1 $\pm$ 0.7 & 1740 $\pm$ 50  & 58.6 $\pm$ 3.7   & 5879.1 $\pm$ 0.5 &   420  $\pm$ 70   &   2.3 $\pm$ 0.5 & 6563.2 $\pm$ 0.8 & 1970 $\pm$ 50 	& 154.7	 $\pm$ 8.1   \\

\hline
\end{tabular}%
}

\textit{Note.} With $\cdots$ we indicate an epoch in which the line in question could not be fit. For the \hei\ $\lambda$5876, the values marked with * are the combination of the values measured for the narrow and broad component. 
\end{table*}
\end{center}

\bibliography{bibliography}{}
\bibliographystyle{aasjournal}

\end{document}